\documentclass[prb,twocolumn,superscriptaddress,preprintnumbers]{revtex4-1}

\usepackage{graphicx}
\usepackage{dcolumn}
\usepackage{bm}
\usepackage{hyperref}
\usepackage{subfigure}
\usepackage{wrapfig}
\usepackage{amssymb,latexsym, amsmath}
\usepackage{graphics}
\usepackage{float}
\usepackage{appendix}
\usepackage{amssymb,latexsym,amsthm,makeidx,enumerate}
\usepackage{mathrsfs}

\usepackage{siunitx}
\usepackage{comment,hhline,multirow,setspace}

\begin{document}

\makeatletter
\let\NAT@bare@aux\NAT@bare
\def\NAT@bare#1(#2){%
	\begingroup\edef\x{\endgroup
		\unexpanded{\NAT@bare@aux#1}(\@firstofone#2)}\x}
\makeatother

\title{Evidence of incoherent carriers associated with resonant impurity levels and their influence on superconductivity in the anomalous superconductor Pb$_{1-x}$Tl$_x$Te}

\author{P. \surname{Giraldo-Gallo}}
\affiliation{Geballe Laboratory for Advanced Materials and Department of Physics, Stanford University, Stanford, CA 94305, USA}
\affiliation{National High Magnetic Field Laboratory, Tallahassee, Florida 32310, USA}
\affiliation{Department of Physics, Universidad de Los Andes, Bogot\'a 111711, Colombia}
\author{P. Walmsley}
\affiliation{Geballe Laboratory for Advanced Materials and Department of Applied Physics, Stanford University, Stanford, CA 94305, USA}
\author{B. Sangiorgio}
\affiliation{Materials Theory, ETH Zurich, Wolfgang-Pauli-Strasse 27, CH-8093 Z\"urich, Switzerland}
\author{S. C. Riggs}
\affiliation{National High Magnetic Field Laboratory, Tallahassee, Florida 32310, USA}
\author{R. D. McDonald}
\affiliation{Los Alamos National Laboratory, Los Alamos, NM 87545, USA}
\author{L. Buchauer}
\affiliation{LPEM (UPMC-CNRS), Ecole Superieure de Physique et de Chimie Industrielles, Rue Vauquelin, 75005 Paris, France}
\author{B. Fauqu\'e}
\affiliation{LPEM (UPMC-CNRS), Ecole Superieure de Physique et de Chimie Industrielles, Rue Vauquelin, 75005 Paris, France}
\author{Chang Liu}
\affiliation{Ames Laboratory and Department of Physics and
Astronomy, Iowa State University, Ames, Iowa 50011, USA}
\author{N. A. Spaldin}
\affiliation{Materials Theory, ETH Zurich, Wolfgang-Pauli-Strasse 27, CH-8093 Z\"urich, Switzerland}
\author{A. Kaminski}
\affiliation{Ames Laboratory and Department of Physics and
Astronomy, Iowa State University, Ames, Iowa 50011, USA}
\author{K. Behnia}
\affiliation{LPEM (UPMC-CNRS), Ecole Superieure de Physique et de Chimie Industrielles, Rue Vauquelin, 75005 Paris, France}
\author{I. R. Fisher}
\affiliation{Geballe Laboratory for Advanced Materials and Department of Applied Physics, Stanford University, Stanford, CA 94305, USA}

\date{\today}

\begin{abstract}

We present a combined experimental and theoretical study of the evolution of the Fermi surface of the anomalous superconductor Pb$_{1-x}$Tl$_x$Te as a function of thallium concentration, drawing on a combination of magnetotransport measurements (Shubnikov de Haas oscillations and Hall coefficient), Angle Resolved Photoemission Spectroscopy (ARPES), and density functional theory (DFT) calculations of the electronic structure. Our results indicate that for Tl concentrations beyond a critical value the Fermi energy coincides with resonant impurity states in Pb$_{1-x}$Tl$_x$Te, and we rule out the presence of an additional valence band maximum at the Fermi energy. Through comparison to non-superconducting Pb$_{1-x}$Na$_x$Te we argue that the presence of these states at the Fermi energy provides the pairing interaction and thus also the anomalously high temperature superconductivity in this material.

\end{abstract}


\maketitle


Pb$_{1-x}$Tl$_x$Te is a hole-doped narrow band-gap semiconductor that exhibits by far the highest known superconducting critical temperature, T$_c$,  of any material of equivalent carrier density \cite{Ueno, Bustarret}. Superconductivity in very low density systems is in itself unexpected from the conventional phonon-mediated BCS theory of superconductivity, and has prompted discussion of a number of exotic pairing mechanisms. In n-doped SrTiO$_3$, for example, non-adiabatic phonon pairing \cite{GorkovPRB16,GorkovPNAS16}, pairing through the exchange of a plasmonic mode \cite{Takada80} or a soft optical phonon mode \cite{Appel69}, pairing from quantum ferroelectric fluctuations \cite{Edge1}, and an enhancement of the density of states and/or pairing potential due to multi-band Fermi surface effects \cite{BehniaSTOPRL14,BinnigSTO,Fernandes13} have been proposed; whereas enhanced electronic correlations from multi-valley Fermi surface effects have been invoked to explain the recently discovered bulk superconductivity in Bismuth \cite{Cohen64,PrakashBi}. The present case of superconductivity in PbTe, however, has the additional curiosity of occurring only when hole-doping is achieved by thallium (Tl) impurities \cite{Nemov1,Ravich}, which suggests that a unique property of the dopant may be key.

\begin{figure}[!t]
	\vspace{-0.3cm}
	\hspace{-0.3cm}
	\centering
	\includegraphics[scale=0.15]{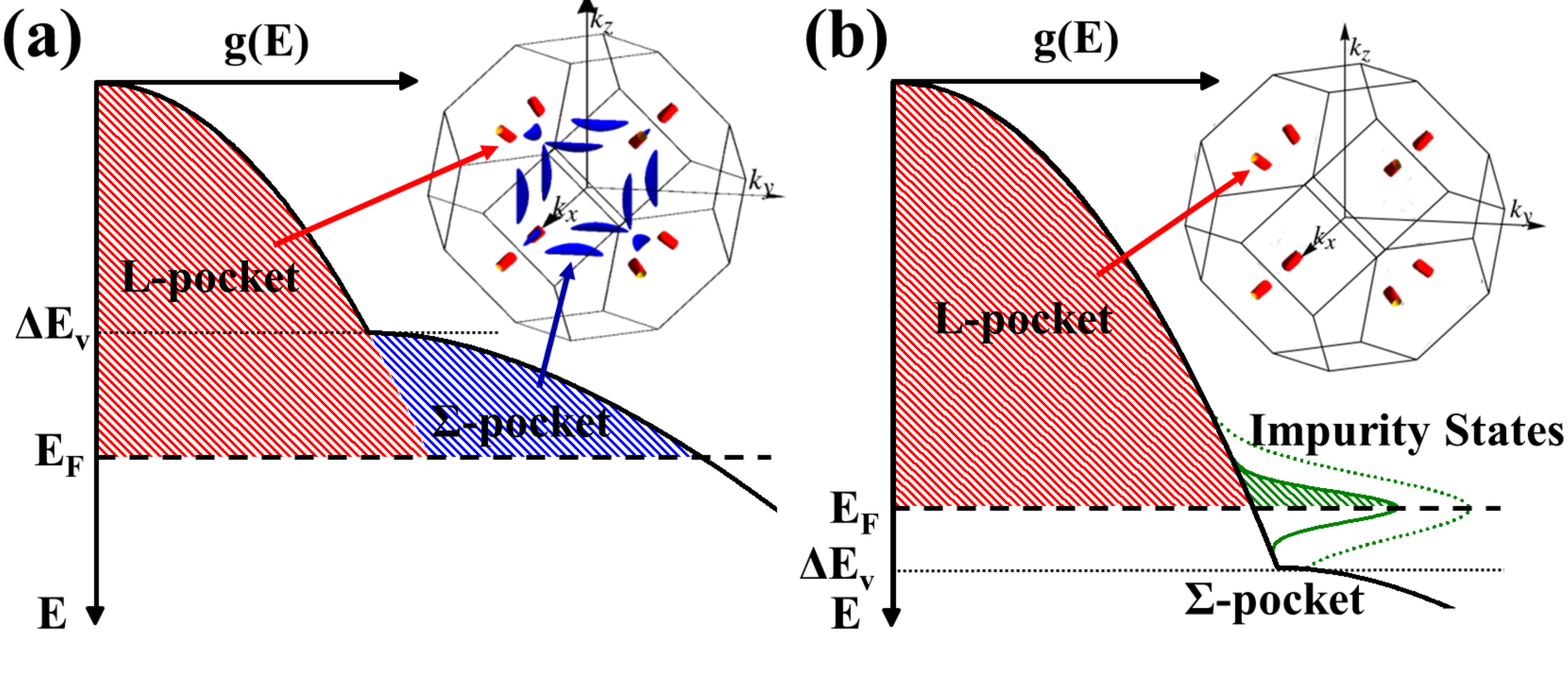} 
	\vspace{-0.6cm}
	\caption{Illustration of two possible scenarios of valence band filling in Pb$_{1-x}$Tl$_x$Te, showing schematic diagrams of the density of states $g(E)$ as a function of energy $E$, and the Fermi surface morphology (insets). {\bf (a)} The situation in which $E_F > \Delta E_v$ and impurity states play no role. The upper valence band maximum (the L-pocket, red) and the second valence band maximum (the $\Sigma$-pocket, blue) are both above the Fermi energy $E_F$, giving four degenerate valleys for L-states, and twelve degenerate valleys for $\Sigma$-states, and an increased density of states for $E_F > \Delta E_v$ where $\Delta E_v$ is the difference in energy between the band maxima. \emph{Inset:} Calculated Fermi surface for $E_F > \Delta E_v$ \cite{GiraldoNaPRB}. {\bf (b)} The situation in which the $\Sigma$-band is never populated because resonant impurity states (shown in green) associated with Tl doping occur at $E=E_F < \Delta E_v$. The Fermi energy $E_F$ is pinned by resonant impurity states at an energy $E_F < \Delta E_v$, leading to an increase in the density of states but only four L-pockets forming the Fermi surface. \emph{Inset:} Calculated Fermi surface for $E_F < \Delta E_v$ \cite{GiraldoNaPRB}. In this paper we establish that scenario (b) is the correct description for Pb$_{1-x}$Tl$_x$Te with $x>x_c$, with immediate implications for the origin of the anomalous superconductivity observed for this material.
	}\label{fig_cartoon}
	\vspace{-0.3cm}
\end{figure}


A number of experimental studies have inferred the presence of shallow impurity levels in the valence band of Pb$_{1-x}$Tl$_x$Te \cite{Nemov1}. Density functional theory (DFT) calculations (appendix A) \cite{Salameh,Cho1}, using supercells containing 64 to 216 atoms, find that low levels of atomic substitution profoundly modify the band structure. Therefore, doping cannot be pictured as a rigid shift of the chemical potential. In the case of Tl, the valence of impurity states depend on the position of Fermi energy, $E_F$. For large (small) hole concentrations, Tl$^{3+}$  (Tl$^{1+}$) is favored \cite{Cho1}. Interestingly, these distinct charge states become energetically degenerate when $E_F$ lies about 50 meV below the top of the valence band, as suggested by earlier heuristic arguments \cite{Nemov1}. 
Charge fluctuations associated with such resonant impurity states have been proposed as an alternative pairing interaction in Tl-doped PbTe \cite{DZero, Matsuura, Costi}, potentially accounting for the anomalously high $T_c$ and other normal state properties \cite{Yana1, Yana2}. However, other scenarios are also possible, in particular a second (heavy) valence band maximum occurs at a similar energy in DFT calculations \cite{GiraldoNaPRB}, raising the alternative view that additional pockets in the Fermi surface could produce a less exotic superconducting mechanism by increasing the density of states and/or providing additional inter-band scattering mechanisms \cite{Ravich}. These two scenarios, illustrated in fig.~\ref{fig_cartoon}, can be distinguished based on characterization of the low-energy electronic structure of Tl-doped PbTe. Here we present a combined experimental and computational study of the low-energy electronic structure and Fermi surface of Pb$_{1-x}$Tl$_x$Te, utilizing Shubnikov de Haas (SdH), angle-resolved photoemission spectroscopy (ARPES), and Hall effect measurements, as well as density functional theory (DFT) calculations of the electronic structure (appendix A). We definitively establish that the $\Sigma$-pocket remains well below the Fermi energy at all compositions of Pb$_{1-x}$Tl$_x$Te and therefore does not drive the superconductivity in this material. Furthermore, by contrasting our data with similar data from the non-superconducting analog Pb$_{1-x}$Na$_x$Te we observe a number of phenomena that distinguish superconducting and non-superconducting compositions, which we show to be consistent with the presence of additional mobile charges associated with the resonant impurity states introduced by Tl doping.

\begin{figure}[!t]
\vspace{-0.2cm}
\hspace{-0.4cm}
\centering
\includegraphics[scale=0.245]{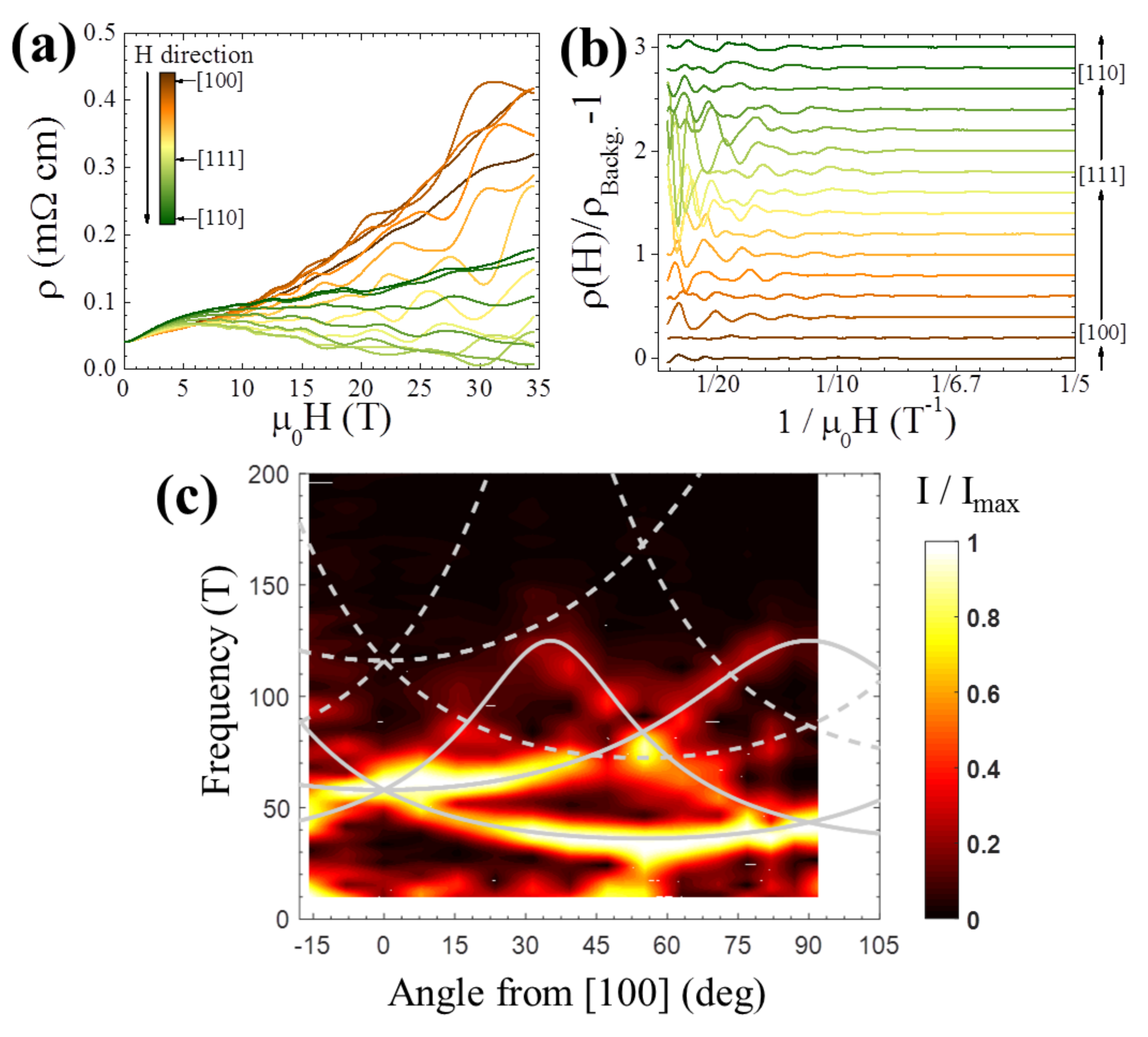} 
\vspace{-0.5cm}
\caption{Magnetoresistance measurements as a function of magnetic field rotated in the (110) plane for a representative sample Pb$_{1-x}$Tl$_x$Te with $x=$0.15$\%$. {\bf (a)} Symmetrized resistivity as a function of applied magnetic field for different field orientations along the (110) plane. {\bf (b)} The oscillating component of the corresponding magnetoresistance curves in (a) as a function of inverse field after the non-oscillating background has been subtracted. {\bf (c)} Amplitude of the normalized FFT, represented by the color scale, as a function of the angle of the magnetic field from the [100] direction, and the frequency. The white curves are the expected angular evolution for a perfect ellipsoidal model (solid-lines for fundamental frequencies, dashed-lines for second-harmonics).
}\label{fig_MR0p15}
\vspace{-0.4cm}
\end{figure}

Single crystals of Pb$_{1-x}$Tl$_x$Te were grown by an unseeded physical vapor transport method (appendix B). For magnetoresistance measurements, the crystals were cleaved into [100] oriented rectilinear bars typically 1 mm in length. Electrical contacts were made in a 4-point geometry along the top face with typical contact resistances of $<1\Omega$ achieved by sputtering gold pads onto lightly scratched surfaces before attaching gold wires with silver epoxy. Compositions $x=0\%$, 0.15$\%$ and 0.4$\%$ were measured to 35 T at the DC facility of the National High Magnetic Field Laboratory (NHMFL) in Tallahassee, FL, USA;  $x=0.3\%$, 0.8$\%$ and 1.3$\%$ were measured to 32 T (DC) at the High Field Magnet Laboratory in Nijmegen, The Netherlands; and additional measurements for $x=0.15\%$ and 0.8$\%$ were taken up to 60 T at the pulsed field facility of the NHMFL, in Los Alamos National Laboratory, NM, USA. Resistivity data was taken at positive and negative fields and symmetrized to obtain the magnetoresistance. The angle dependent data were obtained at a temperature of $(1.5\pm 0.2)$ K using a single-axis rotator probe such that the field direction was rotated through the (110) plane of the crystal. Hall measurements were taken for all the samples studied at a temperature of $1.5$ K. ARPES measurements were performed at beamline 7.0.1 of the ALS and at the SIS beamline of the Swiss Light Source (SLS), Switzerland using a photon energy of $h\nu$=70eV at 40 K. All samples were cleaved in-situ prior to measurement.

\begin{figure}[!t]
\vspace{-0.5cm}
\centering
\hspace{-0.5cm}
\includegraphics[scale=0.25]{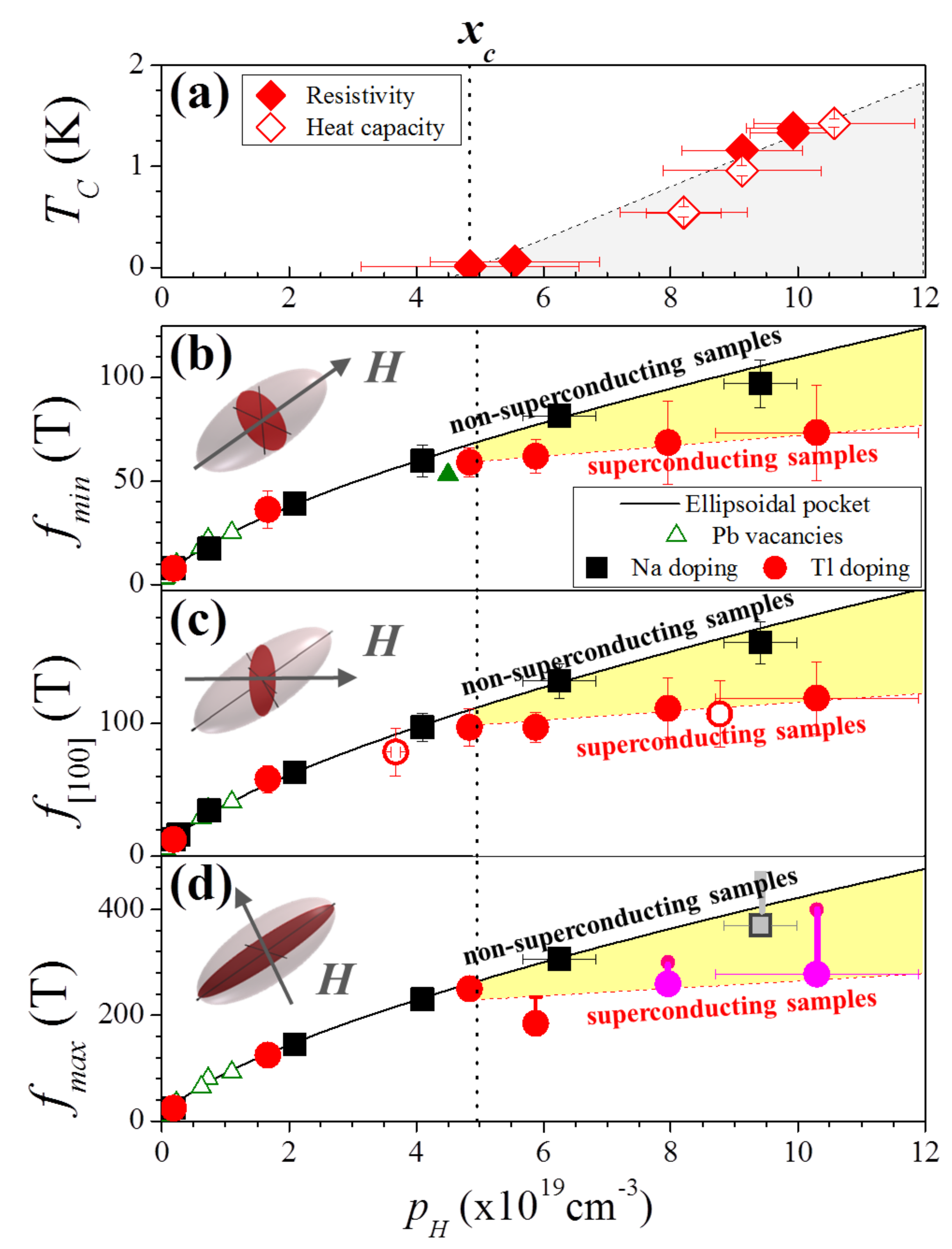}
\vspace{-0.4cm}
\caption{Comparison of $T_c$ (panel (a); from ref.~\citenum{Yana2}) and L-pocket frequencies (panels (b-d)) of Tl-doped PbTe (red data points), non-superconducting Na-doped PbTe (black data points, from ref.~\citenum{GiraldoNaPRB}; and solid-green data point from ref.~\citenum{burke1}), and non-superconducting Pb$_{1-\delta}$Te (open-green data points, from refs.~\citenum{burke1} and \citenum{burke2}). In all four panels the data are plotted as a function of the measured Hall number, and the difference between superconducting and non-superconducting compositions is highlighted by yellow shading. Solid black lines in panels (b-d) illustrate  the functional dependence of $p_H^{2/3}$ expected for a perfect ellipsoidal model with fixed anisotropy. Open red circles represent data obtained from SdH measurements in fields only up to 14 T for Pb$_{1-x}$Tl$_x$Te; the $f_{max}$ data point for the highest Na concentration is represented by a gray square in order to emphasize possible deviations from perfect ellipsoidicity seen in this sample; and the $f_{max}$ data points for the two highest Tl concentrations (pink circles) were determined from $f_{min}$ and assuming a constant pocket anisotropy $K=14.3$ (appendix C). Insets illustrate the field orientations relative to the L-pocket (pink ellipse), corresponding to angles of 54.7$^\circ$, 0$^\circ$ and 35.3$^\circ$ relative to [100]. }
\label{fig_freqsTl}
\end{figure}

Figure~\ref{fig_MR0p15}(a) shows the magnetoresistance of a representative Pb$_{1-x}$Tl$_x$Te sample with $x$=0.15$\%$ ($p_H=1.67\times 10^{19}$cm$^{-3}$) for different field orientations in the (110) plane, clearly exhibiting SdH oscillations. The (110) plane is a natural plane to study the angle evolution of the SdH frequencies in PbTe as it passes through all of the crystallographic high symmetry directions and allows the determination of both the transverse and longitudinal extremal cross-sectional areas of the ellipsoidal pocket known to describe the first valence band maximum situated at the L-point (the `L-pocket') in PbTe. Figure~\ref{fig_MR0p15}(b) shows the oscillating component of the magnetoresistance as a function of inverse field, where the non-oscillating component has been removed by fitting and subtracting a cubic spline from the data in fig.~\ref{fig_MR0p15}(a). Fourier analysis reveals multiple oscillatory components that are periodic in inverse field, as expected for quantum oscillations, with frequencies and magnitudes that vary as a function of angle. This variation is illustrated in fig.~\ref{fig_MR0p15}(c) which shows how the amplitude (represented by the colour scale) of the fast Fourier transform (FFT) of the oscillatory component of the magnetoresistance (fig.~\ref{fig_MR0p15}(b)) evolves with frequency and angle. Similarly to the non-superconducting Na-doped analog \cite{GiraldoNaPRB}, all of the frequencies observed at every angle can be very well fitted by fundamental and second harmonics (shown as solid and dashed lines respectively in fig.~\ref{fig_MR0p15}(c)) corresponding to four ellipsoidal FS pockets at the L-point. This analysis was repeated for each sample (appendix C) and yields the values of the minimum and maximum cross-sectional areas of the L-pocket, which for this particular composition are $f_{min}=(36\pm9)$ T and $f_{max}=(125\pm8)$ T respectively. No signature of the $\Sigma$-pocket or any features of non-ellipsoidicity of the L-pocket were observed at any composition measured in this study.


The values of $f_{min}$, $f_{max}$ and $f_{[100]}$ (SdH frequency with $B\parallel[100]$), of the L-pocket measured here in Pb$_{1-x}$Tl$_x$Te are shown in figs.~\ref{fig_freqsTl}(b-d) as a function of Hall number $p_H$ (which is a good approximation to carrier concentration in a single band-ellipsoidal Fermi surface \cite{Ravichbook,GiraldoNaPRB}), and compared to the same values previously found in non-superconducting Pb$_{1-x}$Na$_x$Te \cite{GiraldoNaPRB} and Pb$_{1-\delta}$Te \cite{burke1,burke2}.  As discussed in ref. \citenum{GiraldoNaPRB}, in Pb$_{1-x}$Na$_x$Te and Pb$_{1-\delta}$Te the cross sectional areas follow the $p_H^{2/3}$ evolution expected for a single-band perfect ellipsoidal model for all the range of carrier concentration (except for the highest Na doping measured, where there may be a small deviation). This model also holds for non-superconducting concentrations of Tl where $x<x_c$, but there is a clear break from this trend for $x>x_c$ in Pb$_{1-x}$Tl$_x$Te with the deviation growing as $x$ increases, concurrently with $T_c$ (see fig.~\ref{fig_freqsTl}(a)). Evidently, the emergence of superconductivity in Pb$_{1-x}$Tl$_x$Te is concomitant with the formation of a second electronic reservoir distinct from the L-pockets. 

\begin{figure}[!t]
\vspace{-0.7cm}
\centering
\hspace{-0.6cm}
\vspace{-0.3cm}
\includegraphics[scale=0.255]{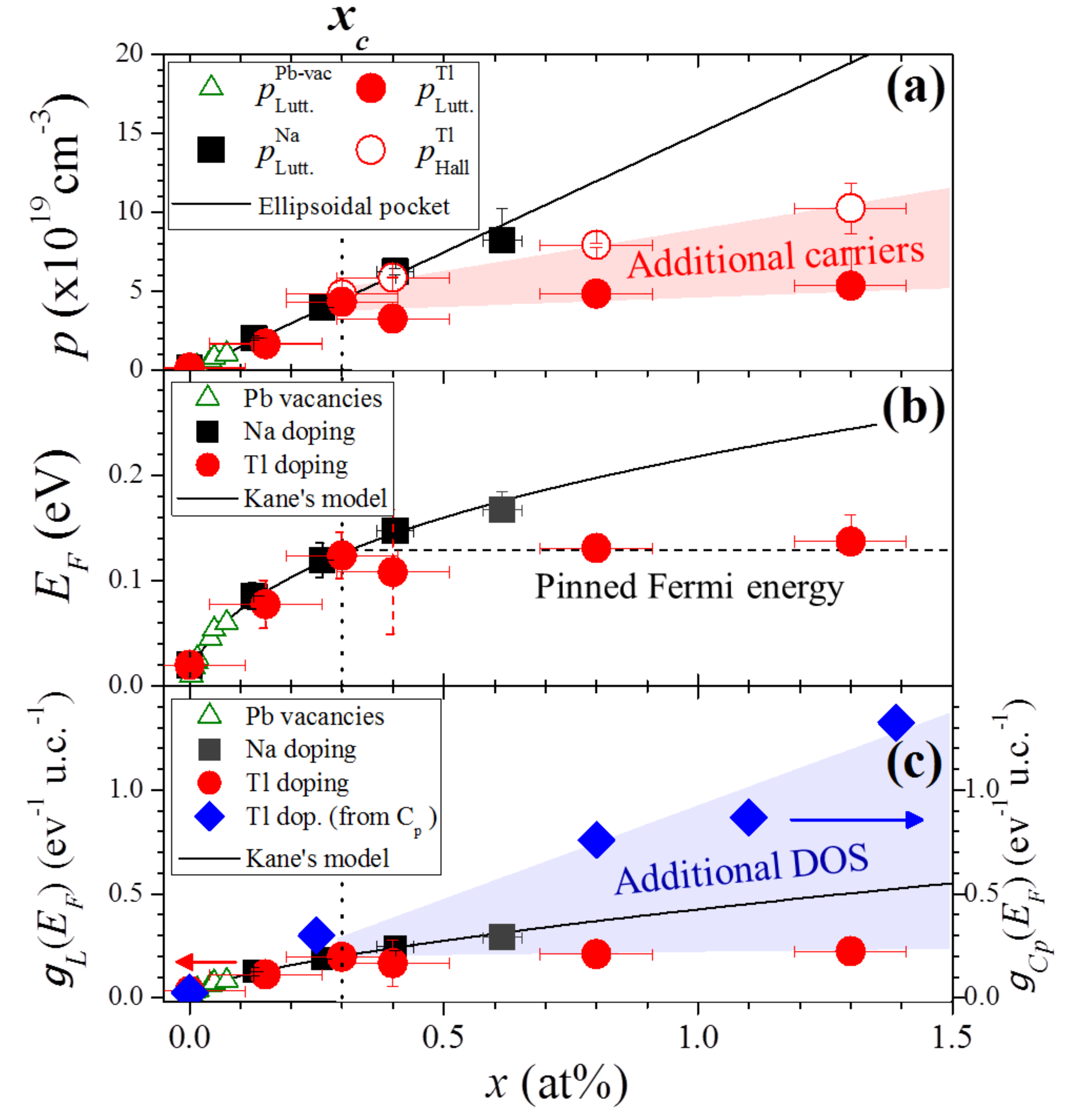} 
\vspace{-0.2cm}
\caption{Evolution of the Luttinger volume of the L-pocket obtained from quantum oscillation measurements (panel (a)), the Fermi energy ($E_F$, panel (b)), and the density of states (DOS) at $E_F$ ($g(E_F)$, panel (c)) as a function of dopant concentration, $x$. Data for self-doped PbTe samples ($p_{\text{Lutt.}}^{\text{Pb-vac.}}$) shown by green-triangles, Na-doped PbTe samples ($p_{\text{Lutt.}}^{\text{Na}}$) by black-squares, and for Tl-doped PbTe samples ($p_{\text{Lutt.}}^{\text{Tl}}$) by red-filled circles. Estimates of the carrier density obtained from the Hall number for Tl-doped samples ($p_{\text{Hall}}^{\text{Tl}}$) are shown by red-open circles in panel (a). In all three panels, the expected evolution of a single band ellipsoidal FS, assuming one hole per dopant and Kane's model, is shown by a black curve \cite{Kane1, Kanebook, GiraldoNaPRB}. (b) Fermi energy and (c) DOS at the Fermi level for doped-PbTe samples as calculated from the experimentally determined L-pocket Luttinger volume and a single-band Kane's model dispersion ($g_L(E_F)$, left-axis). Blue-diamonds in (c) represent the total DOS of Tl-doped PbTe samples as obtained from the electronic contribution to the heat capacity in ref. \citenum{Yana2} ($g_{Cp}(E_F)$, right-axis). Pink-shaded region in panel (a), and blue-shaded region in panel (c), indicate respectively the additional carriers and additional DOS found for Tl-doped PbTe that do not belong to the L-pockets. }
\label{fig_DOSTl}
\vspace{-0.3cm}
\end{figure}

In non-superconducting Pb$_{1-x}$Na$_x$Te the Luttinger volume of the L-pocket and the Hall number are in excellent agreement for all compositions, indicating that all mobile charges are contained within the L-pocket \cite{GiraldoNaPRB}. This is shown in fig.~\ref{fig_DOSTl}(a) where the black line shows the expectation if each dopant contributes a single carrier (or equivalently two carriers per lead vacancy) and the Luttinger volume of Pb$_{1-x}$Na$_x$Te (black squares) follows this closely. However, the present data shows that Pb$_{1-x}$Tl$_x$Te departs from this trend for $x>x_c$. The Hall number (open red circles) falls below the expected value if each Tl dopant were to contribute a single hole (as expected for a Tl$^{1+}$ valence), but still gives a significantly higher estimate of the carrier concentration than the L-pocket Luttinger volume, for composition above $x_c$. The almost unchanging Luttinger volume indicates that the Fermi energy has become almost fixed at around 130 meV for $x>x_c$, in sharp contrast to Pb$_{1-x}$Na$_x$Te (fig.~\ref{fig_DOSTl}(b)). 
The Fermi energy was calculated via Kane's model for an ellipsoidal non-parabolic band that has been shown to be appropriate for the L-pocket in PbTe \cite{GiraldoNaPRB, Ravichbook}. The applicability of Kane's model further allows the electronic density of states of the L-pocket to be estimated and thus compared to the value derived from the Sommerfeld coefficient of the specific heat that necessarily encompasses all states at the Fermi energy. Similarly to the comparison between the Luttinger volume and the Hall effect, if all states are accounted for in the L-pocket then these values should be in agreement, but fig.~\ref{fig_DOSTl}(c) shows a large surplus (blue-shaded area, fig.~\ref{fig_DOSTl}(c)) of density of states observed via heat capacity for $x>x_c$ in Pb$_{1-x}$Tl$_x$Te that cannot be accounted for by the L-pocket observed in the present data. Whilst some of this anomalous increase in the density of states could be due to a zero-bias peak associated with the charge-kondo effect proposed in this material \cite{Yana1}, when taken together with the discrepancy between Luttinger volume and Hall number (shaded area, fig.~\ref{fig_DOSTl}(a)) it is natural to conclude that there must be additional mobile states outside of the L-pocket that continue to increase in number with increasing composition despite the Fermi energy being almost constant.

\begin{figure}[t]
\vspace{-0.7cm}
\centering
\includegraphics[scale=0.45]{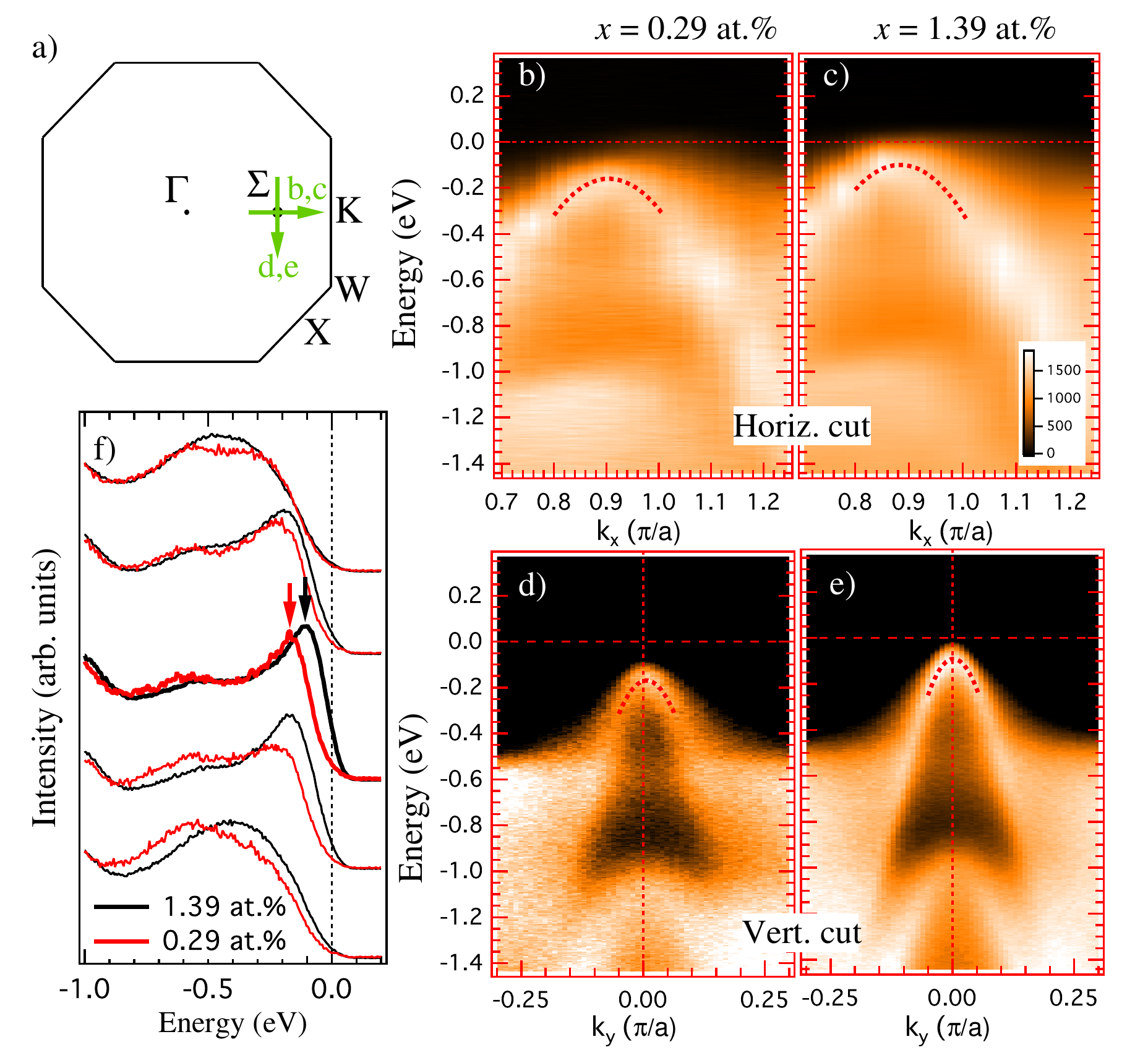} 
\vspace{-0.1cm}
\caption{Band structure analysis of the
$\Sigma$-pocket for two different Tl concentrations illustrating that the $\Sigma$-pocket remains below $E_F$ for all Tl compositions. Dotted curves are guides to the eye. (a) Projection of the Brillouin zone showing the momentum locations of band dispersion data (green arrows). (b)-(c)
Band dispersion data of the $\Sigma$-band along the $\Gamma$-$\Sigma$ direction,
measured for Tl concentration (b) $x = 0.29\%$ and (c) $x =
1.39\%$. (d)-(e) Same as (b)-(c) but along the direction perpendicular to the $\Gamma$-$\Sigma$ direction. 
(f) Energy distribution curves at selected momentum points in vicinity of
the $\Sigma$-pocket top, shown for both Tl dopings. Black and
red arrows point the locations of the top of the $\Sigma$-band.
The top of the $\Sigma$ pocket is found to lie well below $E_F$ for all Tl concentrations.}\label{fig_arpes}
\vspace{-0.3cm}
\end{figure}

Although the data shown in figures~\ref{fig_freqsTl} and \ref{fig_DOSTl} confirm the presence of additional carriers outside of the L-pocket, they do not unambiguously differentiate between the resonant impurity state and $\Sigma$-pocket scenarios. The quantum oscillation amplitude is expected to be significantly smaller for the $\Sigma$-pocket owing to its low mobility relative to the L-pocket \cite{Ravichbook} and therefore could be present but unobserved in the data. Furthermore, the relatively large density of states expected for the $\Sigma$-pocket could give the impression that the Fermi energy is pinned, which is the expectation for an impurity state that is well defined in energy, as $\frac{dE_F}{dx}$ is significantly reduced as $E_F$ reaches the $\Sigma$-band edge. To confirm that the $\Sigma$-pocket remains below $E_F$ for $x>x_c$ in superconducting Pb$_{1-x}$Tl$_x$Te low-temperature ARPES measurements were performed, the results of which are shown in figure~\ref{fig_arpes}. Fig.~\ref{fig_arpes}(a) shows the location and direction in the $k_x$, $k_y$ plane of the band dispersion data shown in figs.~\ref{fig_arpes}(b-e) for the main and transverse axes of the $\Sigma$-pocket. In both compositions measured, $x$=0.29\%$\approx x_c$ (fig.~\ref{fig_arpes}(b\&d)) and $x$=1.39\%$>x_c$ (fig.~\ref{fig_arpes}(c\&e)), the $\Sigma$ pocket is seen to remain below $E_F$ by approximately 170meV and 100meV respectively. This is consistent with existing ARPES results that also conclude that only the L-pocket crosses $E_F$ for a sample of $x$=0.5\% \cite{Nakayama}. This result strongly supports the conclusion that the $\Sigma$-pocket is not responsible for the anomalous fermiology and superconductivity in Pb$_{1-x}$Tl$_x$Te at $x>x_c$. 


The experimental results presented here lead to a number of robust conclusions about the nature of the Fermiology of Pb$_{1-x}$Tl$_x$Te. The key conclusion is that the $\Sigma$-pocket remains well below $E_F$ at all compositions measured of Pb$_{1-x}$Tl$_x$Te, directly showing that this additional band maximum is not responsible for superconductivity. Moreover, $E_F$ becomes pinned in a narrow band of energies at $x>x_c$, coincidently with the onset of superconductivity, which has previously been inferred but never definitively established \cite{Nemov1, Ravich, Yana1}. Also, despite $E_F$ being pinned, we find evidence that additional mobile carriers are present but not associated with a coherent Fermi surface pocket, which could not previously have been inferred from existing data \footnote{The absence of the impurity band in our ARPES measurements is explained by unfolding our density functional calculations of Tl-doped PbTe supercells (see SI), which show that the band has low spectral weight.}. All of these conclusions strongly support the impurity band scenario illustrated in fig.~\ref{fig_cartoon}(b), and are inconsistent with the $\Sigma$-pocket scenario shown in fig.~\ref{fig_cartoon}(a). This study provides the strongest experimental evidence to date that attention should be focused on the Tl impurity states in order to understand the superconductivity of Tl-doped PbTe. 


It remains to be seen how the resonant impurity states contribute to the enhanced superconductivity in Tl-doped PbTe. Existing data in very heavily doped Pb$_{1-x}$Na$_x$Te achieves a similar density of states to that seen in superconducting Pb$_{1-x}$Tl$_x$Te yet still does not exhibit superconductivity down to 10 mK, which suggests that it is the pairing interaction ($V_{kk'}$ from the BCS gap equation) that is introduced extrinsically by the Tl dopants \cite{Chernik1981} that is the key factor determining the onset of superconductivity. Phenomenological models that capture the elements of the electronic structure associated with these impurities certainly indicate that charge fluctuations can provide an effective pairing interaction \cite{DZero,Varma1988}. It is also possible that such correlation effects can lead to an enhanced e-ph interaction, similar to what has been proposed for doped BaBiO$_3$ \cite{KotliarPRX2013}. Independently of these questions, the understanding of the evolution of the electronic structure of Pb$_{1-x}$Tl$_x$Te points the way towards the design of other superconductors based on a similar mechanism.

We thank Y. Matsushita and A. S. Erickson for crystal growth of some of the samples used in this study, T. Geballe, M. Fechner and P. V. C. Medeiros for insightful discussions, and C. Bell for insights that have improved the manuscript. PGG, PW, and IRF, and work performed at Stanford University, were supported by AFOSR Grant No. FA9550-09-1-0583. BS and NAS acknowledge support from ETH Z\"urich, the ERC Advanced Grant program (No. 291151), and the Swiss National Supercomputing Centre (CSCS) under project IDs s307, s624 and p504. The high-field magnetoresistance measurements were performed at the National High Magnetic Field Laboratory (NHMFL), which is supported by NSF DMR-1157490, the State of Florida (DC facility), and the DOE/BES “Science at 100T” grant (pulsed field facility); and the  High Field Magnet Laboratory, Radboud University/Fundamental Research on Matter, member of the European Magnetic Field Laboratory and supported by EuroMagNET II under European Union (EU) Contract 228043. Work at Ames Laboratory (ARPES measurements) was supported by the U.S. Department of Energy, Office of Basic Energy Sciences, Division of Materials Sciences and Engineering. Ames Laboratory is operated for the U.S. Department of Energy by the Iowa State University under Contract No. DE-AC02-07CH11358.  The Advanced Light Source is supported by the Director, Office of Science, Office of Basic Energy Sciences, of the U.S. Department of Energy under Contract No. DE-AC02-05CH11231. 


%



\vspace{2cm}

\appendix
\section*{Appendix}
\setcounter{figure}{0} 
\setcounter{table}{0} 
\makeatletter 
\renewcommand{\thefigure}{A\@arabic\c@figure}
\renewcommand{\thetable}{A\@Roman\c@table}
\renewcommand{\thesubsection}{A.\Roman{subsection}}
\makeatother

\section{First-principles calculations}

In this section we present our density functional calculations.

\subsection{Computational details}

Our first-principles calculations were performed using the PAW~\cite{bloechl1994,kresse1999} implementation of density functional theory (DFT) within the VASP package~\cite{kresse1996}. 
We used a $3\times3\times3$ cubic supercell (216 atoms) and substituted two lead atoms with thallium giving a thallium concentration of $x\approx1.9\,\%$. 
All results were averaged over the nine symmetry-inequivalent configurations resulting from this supercell size. 
We used the PBEsol~\cite{perdew2008} exchange-correlation functional, a $5\times5\times5$ $\Gamma$-centered $k$-point mesh and a plane-wave energy cutoff of $600$ eV. 
Spin-orbit coupling was not included because of computational cost. 
We used valence electron configurations $5d^{10}6s^26p^2$ for lead, $5s^25p^4$ for tellurium, and $5d^{10}6s^26p^1$ for thallium. 
The unit cell volume was fixed to the equilibrium volume (lattice constant of $6.44$~\AA\ to be compared with the experimental lattice constant of $6.43$~\AA~\cite{bozin2010}) of the pristine compound obtained with a full structural relaxation. 
The band unfolding was performed using the BandUP code~\cite{medeiros2014,medeiros2015}.

\subsection{Unfolding and effective band structures}

\begin{figure*}[!ht]
	\centering
	\includegraphics[width=.8\textwidth]{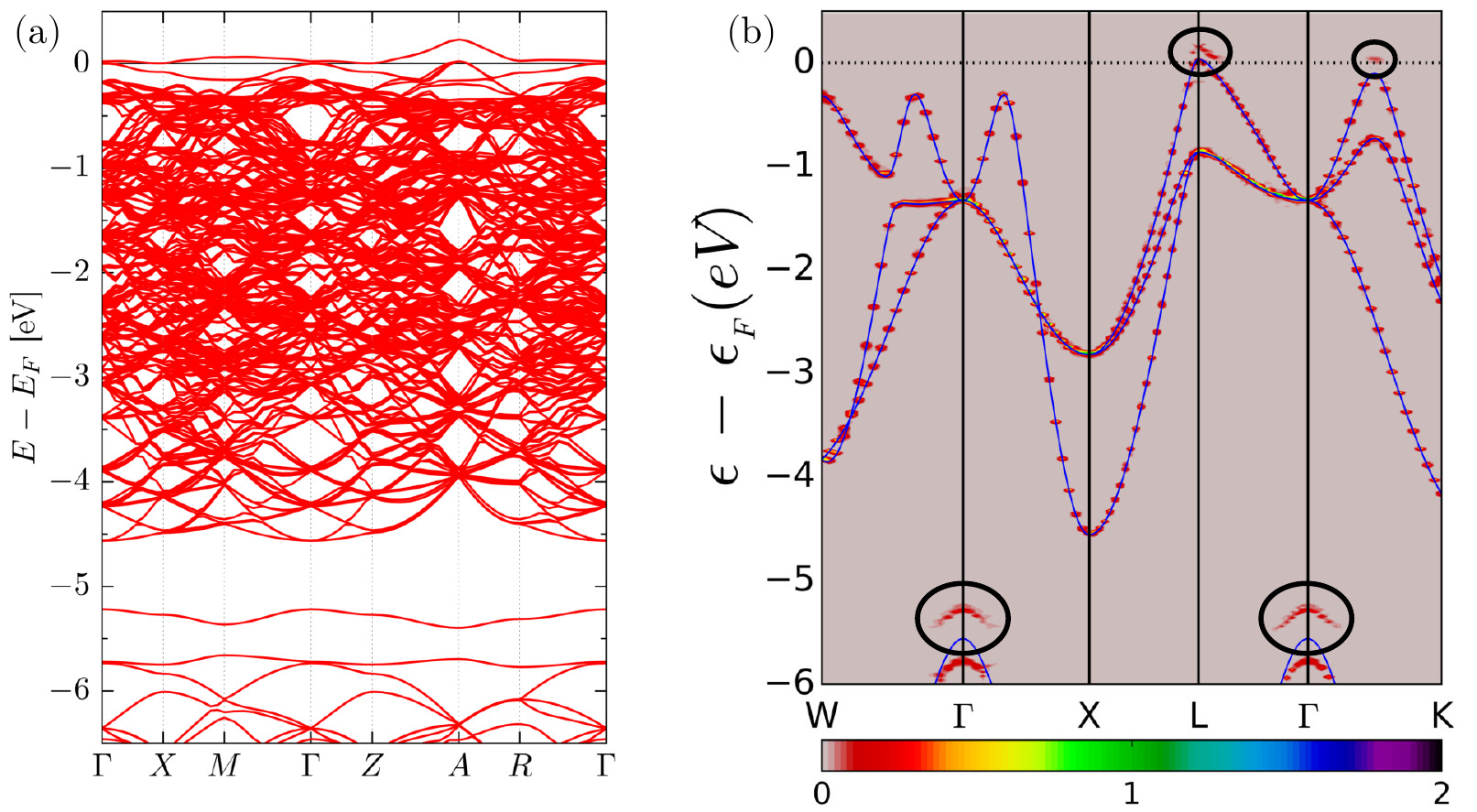}
	\caption{Computed band structures of Tl-doped PbTe showing the valence bands. 
	(a) Band structure in the supercell representation. The anti-bonding band is visible at the Fermi level (see Fig.~\ref{fig:bands-ef}(a) for more details), and the bonding states at about $-5.3$ eV. 
	(b) Effective band structure obtained from unfolding the supercell band structure. The blue lines represent the band structure of pristine PbTe shifted up such that the top of the Te $p$ valence bands match. 
	The black circles indicate where the two impurity bands contribute appreciable spectral weight (see Fig.~\ref{fig:bands-ef}(b) for more details about the anti-bonding band). 
	}
	\label{fig:bands-gen}
\end{figure*}
\begin{figure*}[!ht]
	\centering
	\includegraphics[width=.8\textwidth]{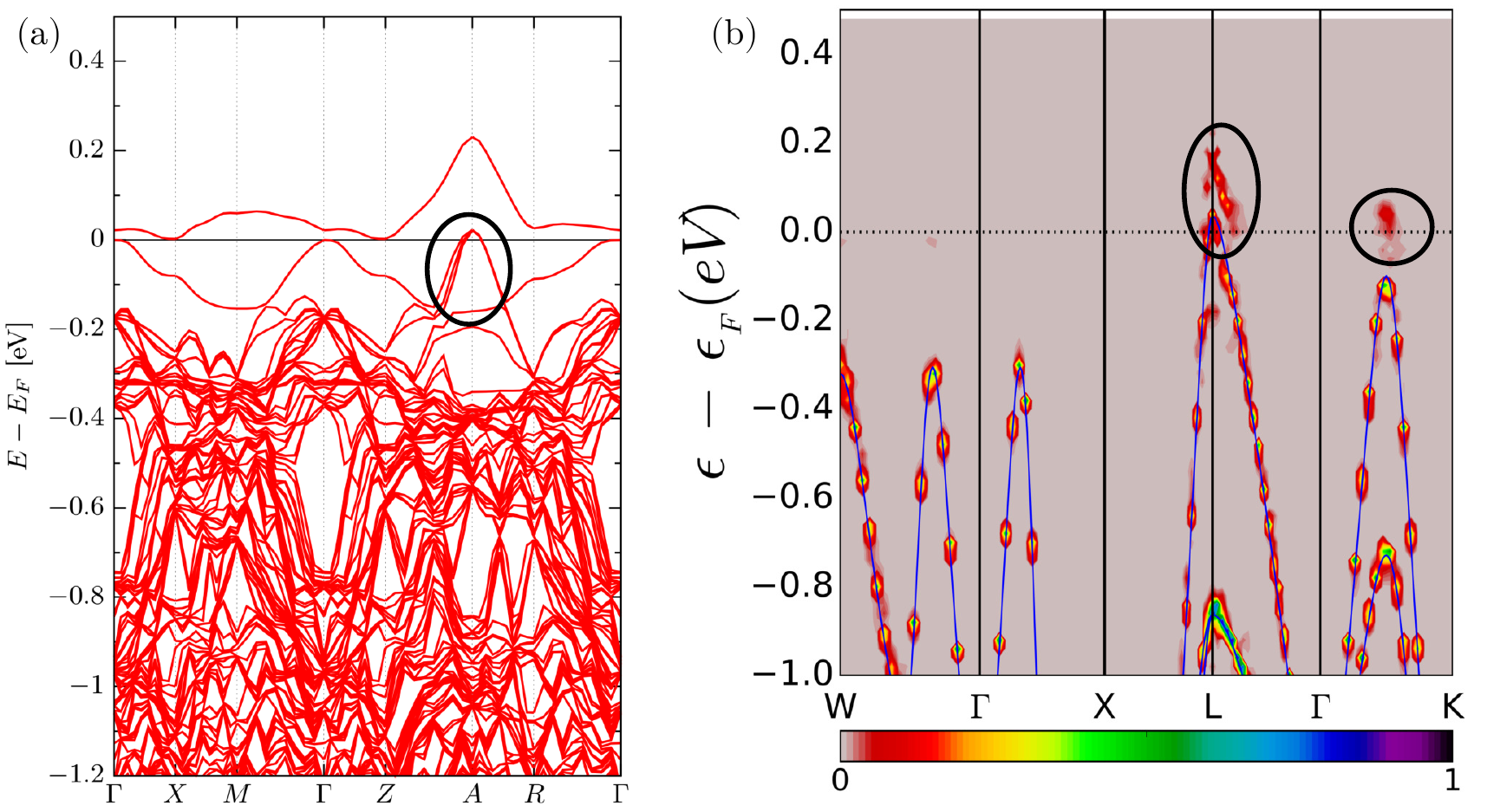}
	\caption{Computed band structures of Tl-doped PbTe close to the Fermi level. 
		(a) Band structure in the supercell representation. At the Fermi level one can observe the narrow  anti-bonding band. 
		The black circle indicates the Te $p$ pocket located at the $L$ point in the primitive-cell description. 
		(b) Effective band structure obtained from unfolding the supercell band structure. The blue lines represent the band structure of pristine PbTe shifted up such that the top of the Te $p$ valence bands match. 
		The black circles indicate where the anti-bonding band contributes appreciable spectral weight. 
	}
	\label{fig:bands-ef}
\end{figure*}
\begin{figure*}[!htb]
	\centering
	\includegraphics[width=.98\textwidth]{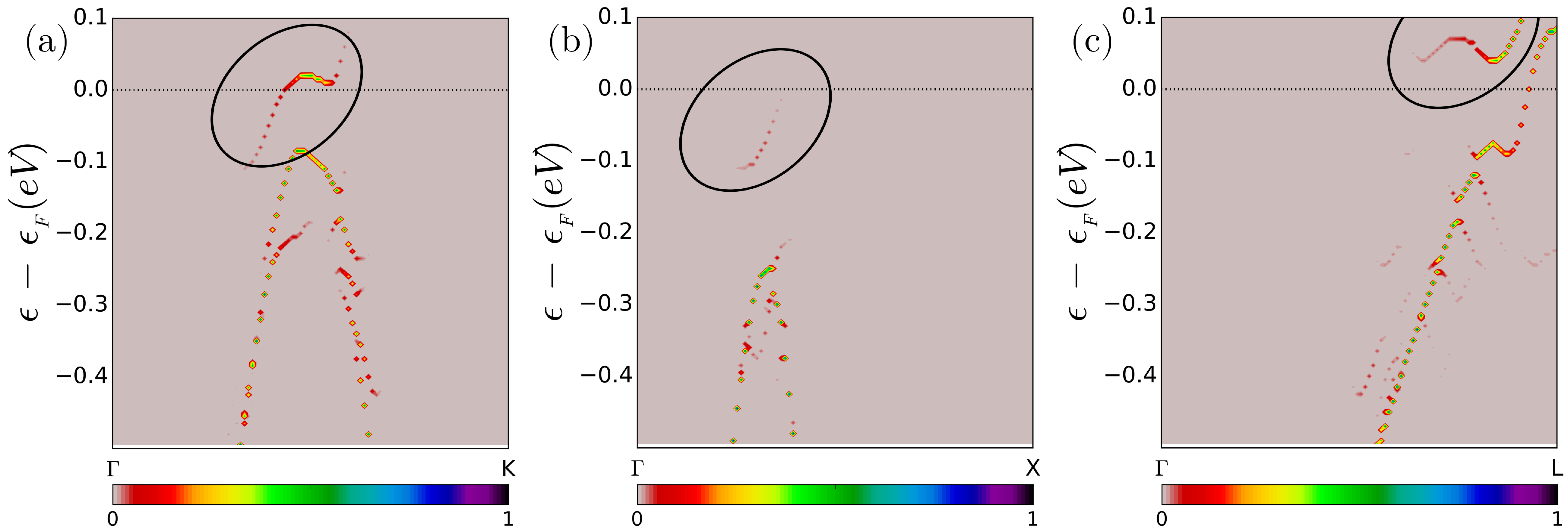}
	\caption{Effective band structures of Tl-doped PbTe close to the Fermi level along three high-symmetry directions, the $\Sigma$ line (a), the $\Delta$ line (b), and the $\Lambda$ line (c). 
	The black circles indicate where the anti-bonding band has nonzero spectral weight. 
	}
	\label{fig:bands-high}
\end{figure*}

Alloys and systems with impurities can be tackled primarily in two different ways from a computational point of view. 
Firstly, using effective medium approaches, like the virtual crystal approximation (VCA) and the coherent potential approximation (CPA). 
In this way one obtains directly the properties in the language of the primitive cell description and a comparison with experiments is straightforward. 
On the other hand, structural relaxations and changes in bond strength and character can not be taken into account. 
Secondly, using supercells with the explicit insertion/substitution of impurity atoms. 
In this way any possible effect and interaction caused by the impurities is automatically taken into account. 
The unfolding technique then allows one to construct an effective band structure (EBS) 
with the primitive-cell translational symmetries allowing a direct comparison with experiments. 

In this work we follow the theory presented in Refs.~\cite{Wang1998,Popescu2010,Popescu2012,medeiros2014,medeiros2015}; for more details about the meaning and interpretation of the unfolding technique and the computational results presented in this Supplementary see also Ref.~\cite{sangiorgio2017}. 
The idea is to calculate the eigenstates of the supercell and determine their amount of primitive-cell Bloch character. 
In other words, one determines the amount of each Bloch state of the primitive cell contained in the supercell wavefunction by decomposing the supercell eigenstates into Bloch states of the primitive cell. 
This is done by computing a spectral function $A(\vec k,\, E)$ associated with each wavevector $\vec k$ in the primitive-cell Brillouin zone. 
In the case of weak symmetry-breaking  disorder this spectral function matches results from ARPES experiments, except for the matrix elements describing the transition from the initial to the final state~\cite{Ku2010,Allen2013}. 
Finally, the EBS can be obtained by assigning a weight $\delta N$ to each point $(\vec k,\, E)$ using the cumulative probability function $\mathrm dS(\vec k,\,E)=A(\vec k,\, E)\,\mathrm dE$~\cite{medeiros2014,medeiros2015}.
$\mathrm dS(\vec k,\, E)$ represents the number of bands crossing the energy interval $(E,\,E+\mathrm dE)$ at the wavevector $\vec k$ in the primitive-cell Brillouin zone. 
 
\subsection{Results}

Figure~\ref{fig:bands-gen} shows the computed valence bands. Panel (a) presents the band structure in the supercell representation; only one representative symmetry-inequivalent configuration is shown. 
As in previous studies~\cite{ahmad2006,ahmad2006a,Hoang2008,Xiong2010} we find that the thallium impurities form two localized states, one at the Fermi level (also called resonant, deep-defect or anti-bonding state) and one at about $-5.3$~eV (also called hyperdeep-defect or bonding state). 
These states originate from hybridization of the Te $p$ and Tl $s$ states. 
Panel (b) presents the EBS obtained from unfolding and averaging the supercell band structures of the nine symmetry-inequivalent configurations. 
One can observe some weak contributions, i.e.\ the associated spectral weight is quite small, from the anti-bonding and bonding states (black circles), but only in certain regions of the Brillouin zone; around the $\Gamma$ point for the bonding state and around the $L$ point and along the $\Sigma$ ($\Gamma-K$) line for the anti-bonding state.   

Figure~\ref{fig:bands-ef} shows a zoom of the band structure of Fig.~\ref{fig:bands-gen} in the region close to the Fermi level. In panel (a) one can nicely observe the narrow anti-bonding band at the Fermi level and the Te $p$ pocket (black circle) located at the $L$ point in the primitive-cell picture. 
Panel (b) presents the EBS in the same energy range. 
The anti-bonding band has appreciable spectral weight only around the $L$ point and along the $\Sigma$ line (black circles). 
From this plot it is not clear whether and where in the Brillouin zone the anti-bonding band crosses the Fermi level. 

For this reason we show in Fig.~\ref{fig:bands-high} the EBS along three high-symmetry directions, the $\Sigma$ line (a), the $\Delta$ line (b), and the $\Lambda$ line (c). Note again that the anti-bonding band has appreciable spectral weight only in some parts of the considered lines (black circles). 
This band can be seen to cross the Fermi energy once along the $\Sigma$ line, and also along the $\Delta$ line, even if in the latter case the spectral weight is so small that it can be hardly seen on the printed plot. 
On the other hand, no crossing is observed along $\Lambda$; the observed crossing originates from the Te $p$ bands. 
The observed trend suggests that the Fermi surface produced by the anti-bonding band would contain the $K$ and $X$ points, but not the $\Gamma$ and $L$ points. 
A more accurate description of the shape of the Fermi surface requires a scan through the region of interest in the Brillouin zone; it was not undertaken because of computational cost. 

In summary, we have shown that the impurity band of Tl-doped PbTe is composed of partly delocalized states since the associated unfolded bands show some dispersion. 
On the other hand, its contributions can not be observed throughout the whole Brillouin zone, but only in certain regions. 
Even in those regions, its spectral weight is small, explaining why these states are undetected in the ARPES measurements shown in this work.

\section{Crystal Growth}

Pb$_{1-x}$Tl$_x$Te single crystals were grown by an unseeded physical vapor transport (VT) method, similar to that described in Ref. \citenum{Yana2}, by sealing in vacuum polycrystalline pieces of the already doped compound, with (or close to) the desired final stoichiometry. The polycrystalline material was obtained by mixing high purity lead telluride (PbTe, 99.999$\%$, Alfa Aesar), Thallium telluride (Tl$_2$Te, 99.9$\%$, Alfa Aesar) and metallic tellurium (99.999+$\%$, Alfa Aesar) in the appropriate ratios. The mix was ground to fine powders, in air, and then cold-pressed into a compact pellet. This pellet was sealed in a quartz tube with a small argon pressure and then sintered at 600 $^{\circ}$C for 24 hours. The grind-press-sinter process was repeated once more. The pellet was broken into small pieces (0.5-3mm), placed in a quartz tube, with a length of 10 cm and fully evacuated. The tube was placed in a horizontal furnace, heated up to 700-750 $^{\circ}$C and held at this temperature for 10 days. Mass transport occurs from the hottest part of the tube (center, where the material is initially placed) to the cooler ends, at which high quality single crystals nucleate thanks to the slow mass transport. Additionally, depending on slight variations in the furnace temperature gradient, some single crystals are observed to form at the top of the initial polycrystalline pile at the center of the tube, which is possible due to a slight vertical temperature gradient due to contact of the quartz tube with the furnace central-thermocouple.

Determination of the Tl concentration was performed by comparing the measured Hall number of any given sample against a previously established calibration curve. The calibration was obtained by performing multiple Hall effect and electron microprobe measurements for samples grown by the same technique, for a wide range of Tl concentrations up to the solubility limit, as described in ref.  \citenum{Yana0}.

\section{Determination of Fermi Surface parameters}

\begin{figure*}[!htbp]
\vspace{-0.8cm}
\centering
\includegraphics[scale=0.265]{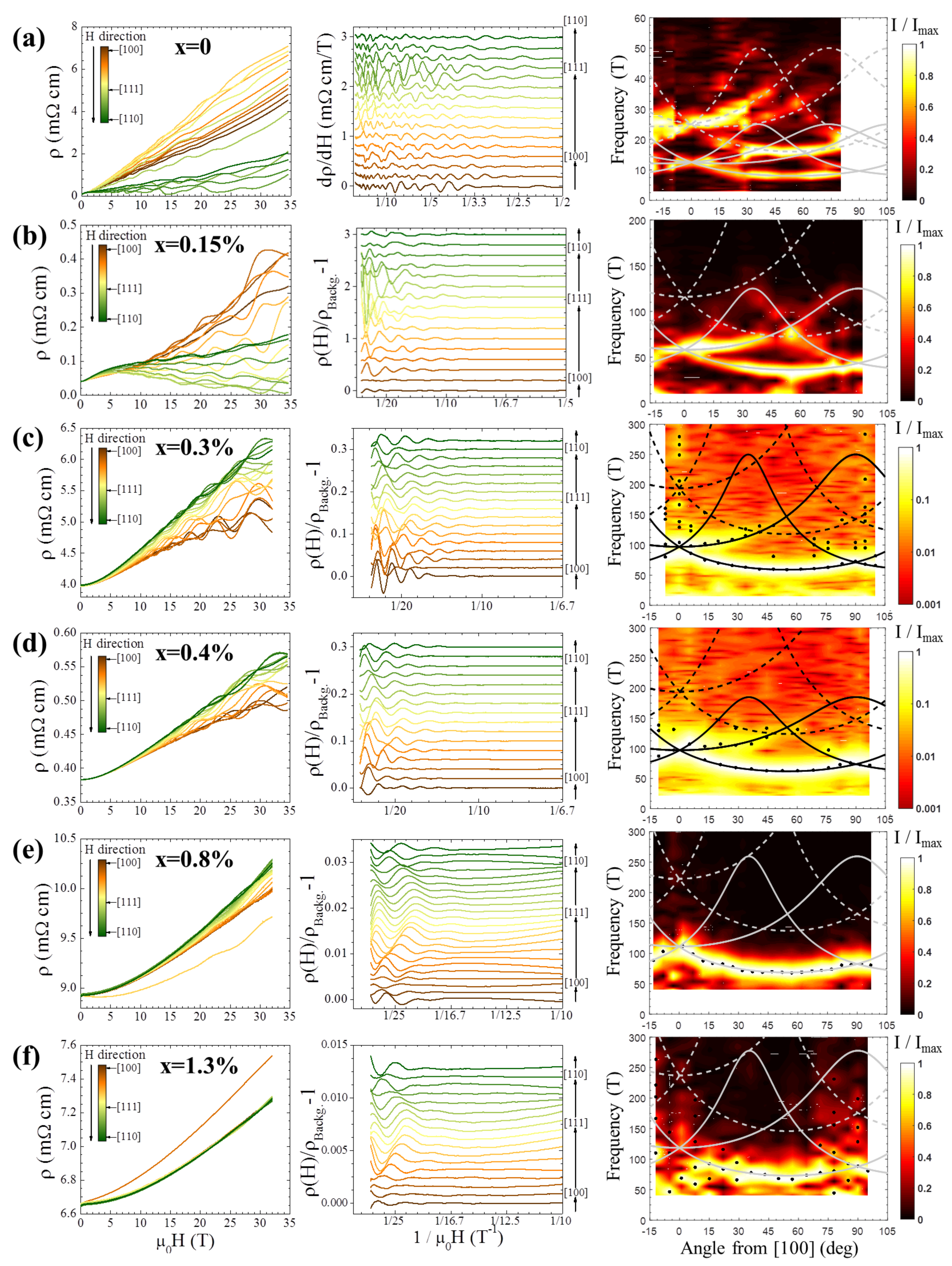} 
\caption{(Color online) Magnetoresistance measurements for Pb$_{1-x}$Tl$_x$Te samples of different Tl concentrations (row \textbf{(a)} $x=$0, row \textbf{(b)} $x=$0.15$\%$, row \textbf{(c)} $x=$0.3$\%$, row \textbf{(d)} $x=$0.4$\%$, row \textbf{(e)} $x=$0.8$\%$ and row \textbf{(e)} $x=$1.3$\%$) as a function of magnetic field rotated along the (110) plane. The first column shows the measured resistivity as a function of applied magnetic field. The second column shows the background-free resistivity, obtained as explained in the text, as a function of inverse field. The third column shows the amplitude of the normalized FFT, represented by the color scale, as a function of the angle of the magnetic field from the [100] direction (horizontal axis), and the frequency (vertical axis). These plots include a comparison to a perfect ellipsoidal model calculation superimposed (solid-lines for fundamental frequencies, and dashed-lines for higher-harmonics). The parameters used for the perfect ellipsoidal model calculation for each set of data are summarized in Table \ref{table_Tlparam}.}
\label{fig_FFTall}
\end{figure*}

\begin{table*}[!htbp]
\centering
\caption{Fermi surface parameters for Tl-doped PbTe, obtained from comparison between data and perfect ellipsoidal model. $p_H$ is the Hall coefficient obtained though Hall measurements (at $T=$1.5 K); $f_\text{min}$ and $f_\text{max}$ are the transverse and longitudinal cross-sectional areas of the L-pockets, respectively; $f_{[100]}$ is the L-pocket cross sectional area perpendicular to the [100] direction; $K=\left(f_\text{max}/f_\text{min}\right)^2$ is the anisotropy of L-pockets; $p_{FS-Vol}$ is the carrier concentration computed using Luttinger's theorem; $m^{cyc}_{[100]}/m_e$ is the cyclotron effective mass along the [100] direction, obtained from the temperature dependence of the amplitude of the oscillating component of magnetoresistance/PDO frequency shift (see section \ref{massSI}). The $f_{max}$ values marked with $*$ were deduced from $f_{min}$ and assuming the same $K$ value ($K=14.3\pm 0.4$) obtained for Na-doped PbTe samples in the same range of carrier concentrations \cite{GiraldoNaPRB}.}\label{table_Tlparam}
\begin{ruledtabular}
\begin{tabular}{ c c c c c c c c }
	$x$(at.$\%$) & $p_H$($\times 10^{19}$cm$^{-3}$) & $f_{min}$ (T) & $f_{[100]}$ (T) & $f_{max}$ (T) & $K$ & $p_{FS-Vol}$ ($\times 10^{19}$cm$^{-3}$) & $m^{cyc}_{[100]}/m_e$  \tabularnewline\hline
	0 & 0.19 $\pm$ 0.0007 & 8 $\pm$ 1 & 12.5 $\pm$ 2 & 25 $\pm$ 2 & 10 $\pm$ 4 & 0.16 $\pm$ 0.02 & - \tabularnewline
	0.15 & 1.67 $\pm$ 0.005 & 36 $\pm$ 9 & 58 $\pm$ 10 & 125 $\pm$ 8 & 12 $\pm$ 6 & 1.7 $\pm$ 0.2 & 0.058 $\pm$ 0.003 \tabularnewline
	0.2 & 3.68 $\pm$ 0.07 & - & 78 $\pm$ 18 &  - &  - & - & - \tabularnewline
	0.3 & 4.84 $\pm$ 0.11 & 59 $\pm$ 7 & 97 $\pm$ 14 & 250 $\pm$ 18 & 18 $\pm$ 5 & 4.4 $\pm$ 0.4 & 0.142 $\pm$ 0.004 \tabularnewline
	0.4 & 5.88 $\pm$ 0.04 & 62 $\pm$ 8 & 97 $\pm$ 11 & 185 $\pm$ 50 & 9 $\pm$ 5 & 3.3 $\pm$ 0.9 & - \tabularnewline
	0.8 & 7.96 $\pm$ 0.14 & 69 $\pm$ 20 & 111 $\pm$ 23 & 260 $\pm$ 40 $*$& 14 & 4.9 $\pm$ 1.0 & 0.233 $\pm$ 0.002 \tabularnewline
	1.1 & 8.77 $\pm$ 0.15 & - & 107 $\pm$ 25 &  - &  - & - & - \tabularnewline
	1.3 & 10.3 $\pm$ 1.6 & 73 $\pm$ 23 & 119 $\pm$ 27 & 278 $\pm$ 122 $*$& 14  & 5.4 $\pm$ 2.5 & - \tabularnewline
\end{tabular}
\end{ruledtabular}
\end{table*}

\begin{figure*}[!htbp]
\centering
\includegraphics[scale=0.24]{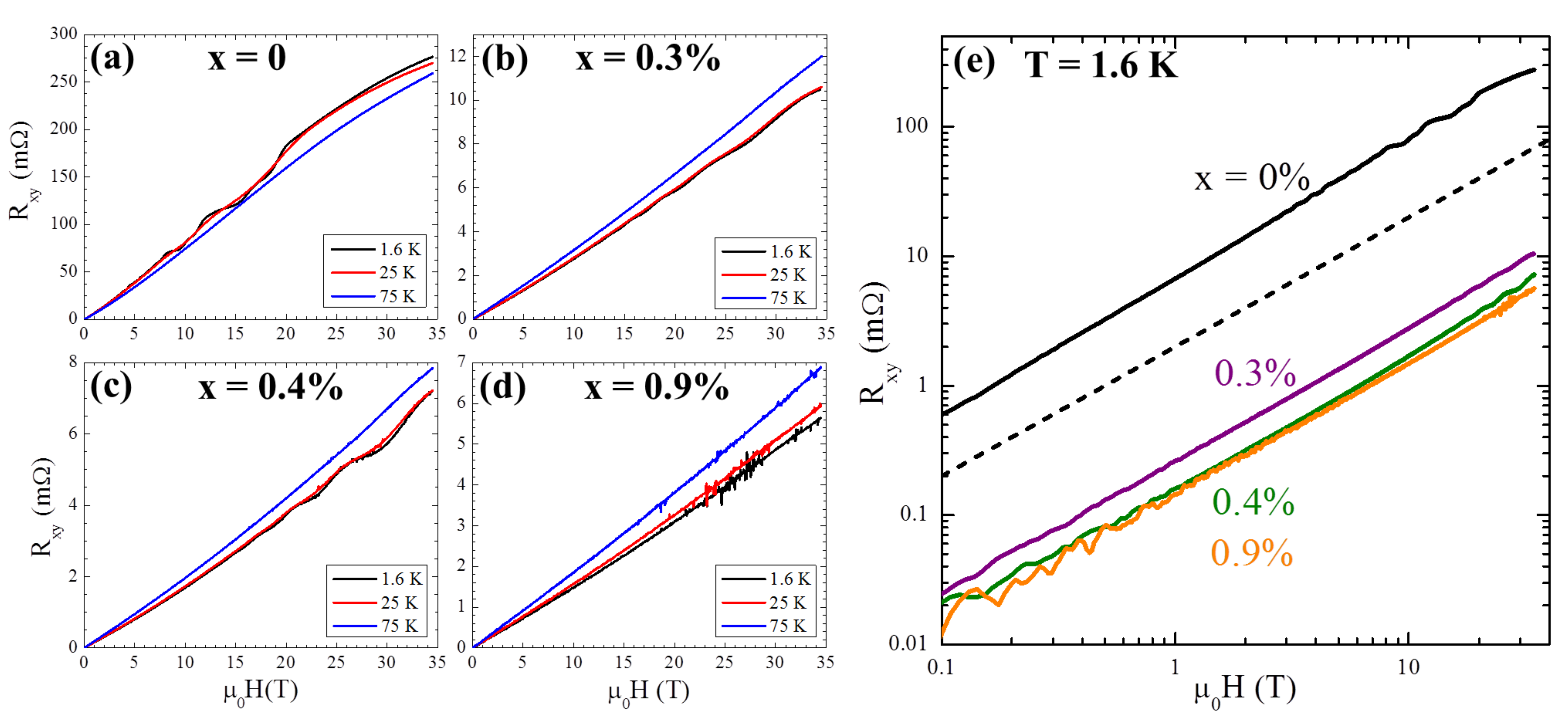} 
\caption{(Color online) Hall resistance $R_{xy}$ for magnetic fields oriented along the [100] direction, for Pb$_{1-x}$Tl$_x$Te samples of different Tl concentrations, and for different temperatures: \textbf{(a)} $x=$0, \textbf{(b)} $x=$0.3$\%$, \textbf{(c)} $x=$0.4$\%$, and \textbf{(d)} $x=$0.9$\%$. \textbf{(e)} Logarithmic plot of the Hall resistance for all the compositions studied, at a temperature of 1.6 K. The dotted line represents a linear functional form, $R_{xy}\propto (\mu_0H)$, expected for a single band parabolic model. The data for all compositions follow this functional form, supporting the single band picture found in our SdH and ARPES results.}
\label{fig_Rxy}
\end{figure*}

The same analysis that was performed for the data shown in Fig.~2 of the main manuscript for the $x$=0.15$\%$ Tl-doped sample was also performed for all the other samples measured. The angle evolution of the frequencies of oscillation for each Tl composition is shown in Figure~\ref{fig_FFTall}. These plots include a comparison with the perfect ellipsoidal model, using the maximum and minimum cross-sectional areas that best fit the data (see Table~\ref{table_Tlparam}). These values are plotted in Fig.~3 of the main manuscript as a function of Hall number. Inspection of the data shown in Fig.~\ref{fig_FFTall} reveals a progressive reduction of the amplitude of oscillations with increasing Tl concentration for compositions above $x$=0.15$\%$, resulting in FFT's with low-amplitude and broad peaks, especially in the higher frequency ranges. As a consequence of this reduction, direct determination of the topology and volume of the L-pockets is more challenging for the highest Tl-concentrations. For the Tl concentrations of $x$=0 (pure PbTe) and $x$=0.15$\%$ the angle evolution of the three different frequency branches for this plane of rotation can be fully tracked for all angles, allowing us to directly obtain the values for the minimum and maximum L-pockets cross-sectional areas from the data. The sample with Tl concentration of $x$=0.3$\%$, shown in Fig.~\ref{fig_FFTall}(c), shows a clear evolution of, at least, two of the frequency branches, which can be tracked up to frequency values of around 150 T. For this sample, a logarithmic color scale was used in order to highlight the weak high-frequency contributions. Additionally, the exact positions of the maxima of the FFT for each angle are shown (black circles), which still permits comparison with the perfect ellipsoidal model. The weak weight seen at the 90$^{\circ}$-250 T region, together with the angle evolution of the middle frequency branch for frequencies below 150 T, provide strong constraints for the determination of the maximum L-pocket cross-sectional area for this composition, even though the dispersion of the whole branch cannot be tracked for all angles. A similar situation is observed for the sample with Tl concentration of $x$=0.4$\%$, shown in Fig.~\ref{fig_FFTall}(d). The color scale in this plot is also logarithmic, and the exact position of the FFT maxima for all angles are shown. For this sample there is no direct indication of the maximum L-pocket cross-sectional area. However, the determination of this quantity was guided by the angle evolution of the middle frequency branch for frequencies below ~130 T, as well as the angle-evolution of the lower-frequency branch, although this last is much less sensitive to variations of $f_{max}$. For the two-highest compositions, just as for all the other compositions, the determination of the L-pocket minimum cross-sectional area, $f_{min}$, is direct. However, the L-pocket maximum cross-sectional area, $f_{max}$, cannot be obtained directly from the FFT contour plots in Figs.~\ref{fig_FFTall}(e) and \ref{fig_FFTall}(f), as only the angle evolution of the lower frequency branch can be observed. For the perfect ellipsoidal model curves shown in these plots, the value of $f_{max}$ was obtained assuming the same $K$ value ($K=14.3\pm 0.4$) found in Na-doped PbTe samples in the same range of carrier concentrations \cite{GiraldoNaPRB}. These values are indicated by asterisks in Table \ref{table_Tlparam}, and are shown by pink symbols in Figure 3(d) of the main text.

\begin{figure}[!ht]
\centering
\includegraphics[scale=0.185]{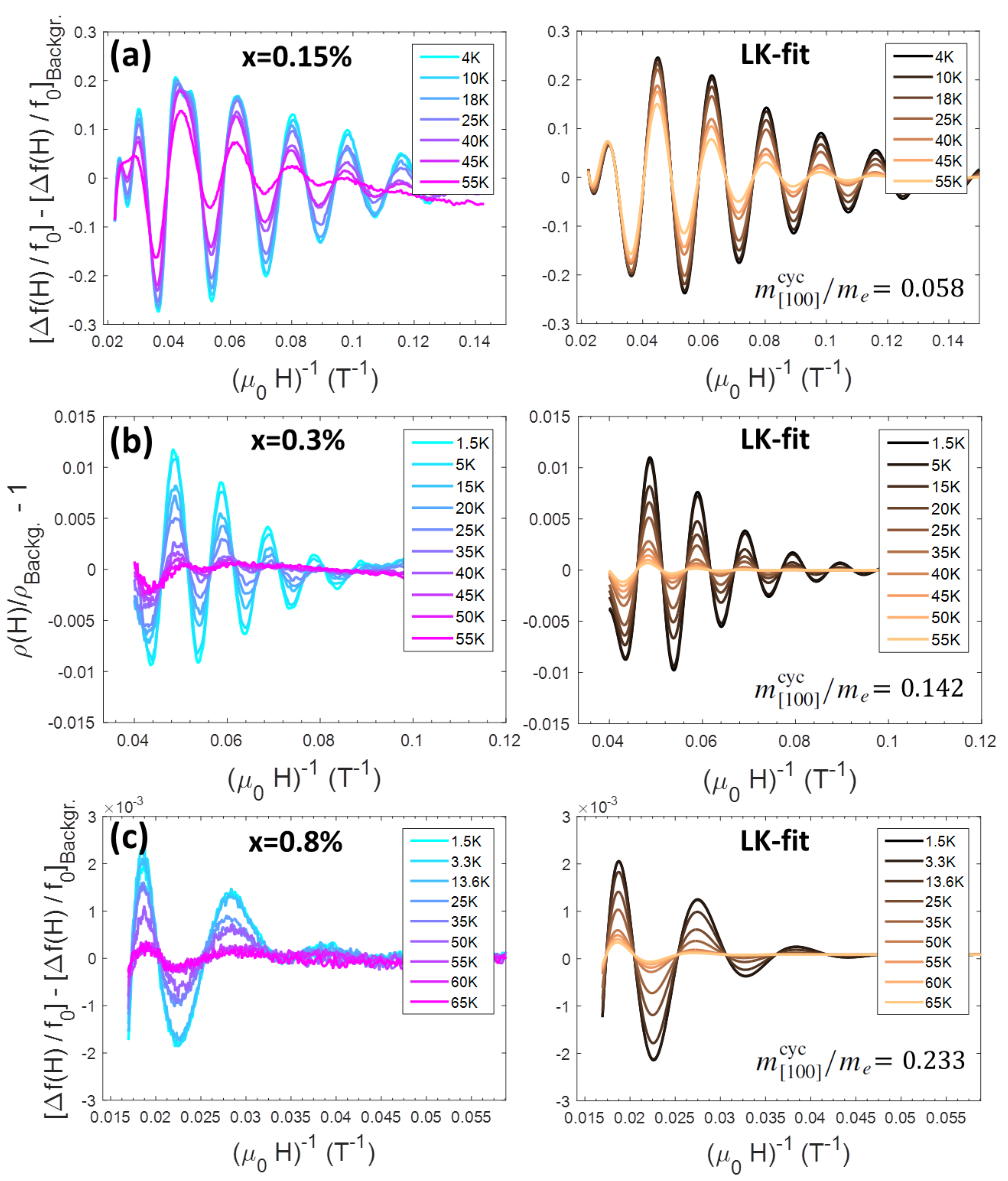} 
\caption{(Color online) Temperature dependence of the amplitude of the oscillating component of magnetoresistance/PDO frequency shift for Pb$_{1-x}$Tl$_x$Te samples, with magnetic field along the [100] direction. The left-column plots show the background-free data at different temperatures. The right-column plots show the fits of the data to the LK-formula, using the two most dominant frequencies observed in the FFT of the lowest temperature curve. From this fit, the values of cyclotron effective mass, for each frequency term, are obtained. The values obtained for the cyclotron mass along the [100] direction, $m_{[100]}^{cyc}/m_e$, for each composition, are shown in the right-column plots and summarized in Table \ref{table_Tlparam}.}
\label{fig_massTl}
\end{figure}

\begin{figure}[!ht]
\centering
\includegraphics[scale=0.3]{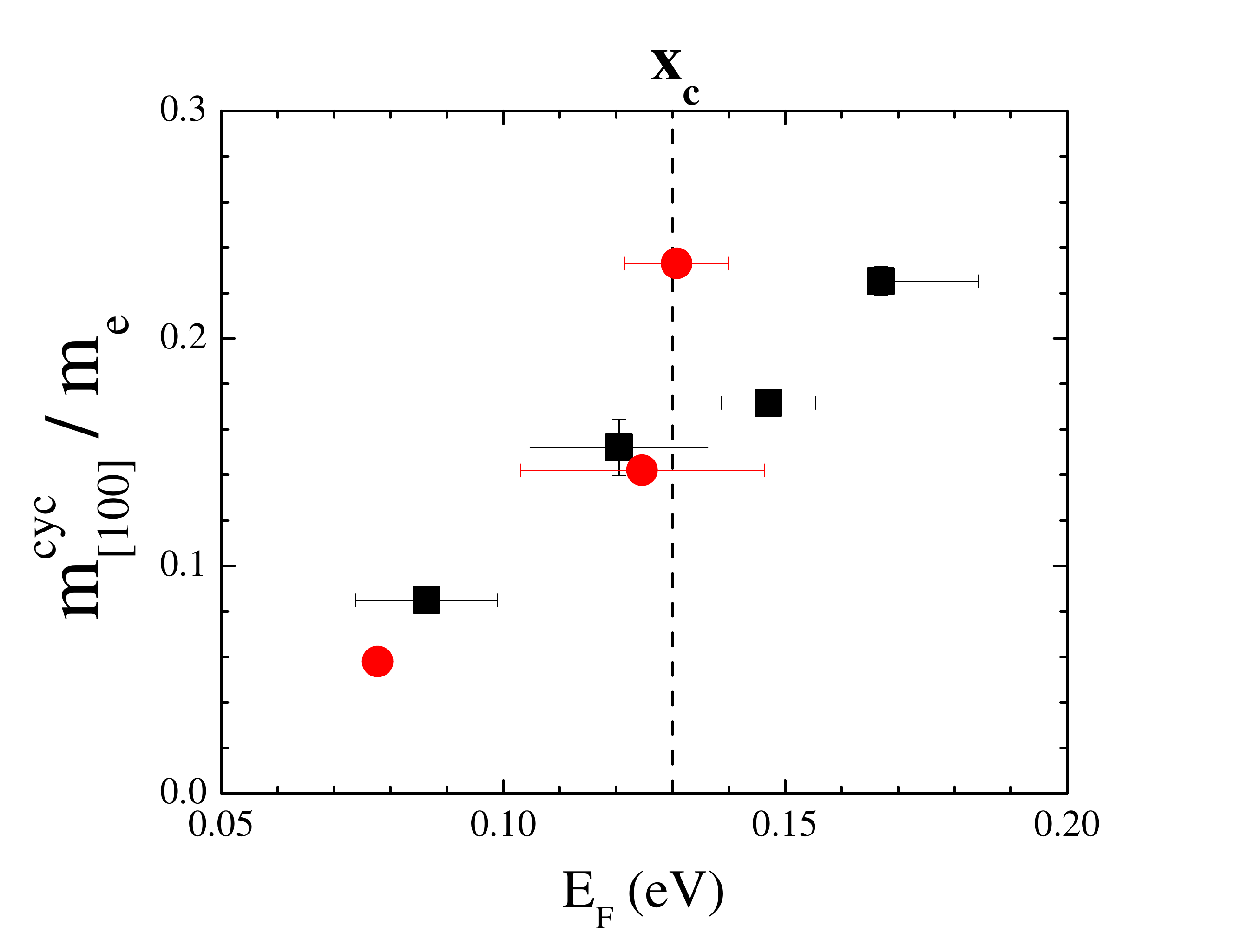} 
\caption{(Color online) Cyclotron effective mass along the [100] direction, $m_{[100]}^{cyc}/m_e$, for Na-doped PbTe samples (black squares, from Ref. \citenum{GiraldoNaPRB}) and Tl-doped PbTe samples (red circles, measured in this work), as a function of Fermi energy. Fermi energies were calculated from the L-pockets Luttinger volume and Kane's dispersion relation \cite{GiraldoNaPRB}. }
\label{fig_masspLutt}
\end{figure}

\section{High field Hall resistance}

Hall effect measurements were performed for representative dopings, for fields up to 35 T at the NHMFL. Fig. \ref{fig_Rxy} shows the data of $R_{xy}$ as a function of magnetic field (oriented in the [001] direction), for temperatures of 1.6 K, 25 K and 75 K, and compositions of $x=0$, $0.3\%$, $0.4\%$ and $0.9\%$. For the lowest Tl concentrations, a small oscillating component is observed in the data, superimposed to the non-oscillating background, which is the focus of these measurements. For the $x=0$ composition deviations from linear behavior, together with a suppression of the oscillatory component, are clearly observed for fields above $~20$ T. These deviations are ascribed to overcoming the quantum limit for this sample at this field orientation. For all other compositions, which are well below the quantum limit for all fields measured, the Hall resistance follows a linear dependence with magnetic field for all fields, supporting the single band picture deduced from SdH and ARPES data. The linear Hall behavior is better observed in Fig. \ref{fig_Rxy}(e), which shows the Hall resistance as a function of field for all compositions at a temperature of 1.6 K, in a logarithmic scale. As can be readily seen, the Hall resistance is a linear function of field for all compositions.

\section{Effective cyclotron mass}\label{massSI}

In order to determine the evolution of effective cyclotron masses of holes in Tl-doped PbTe with carrier concentration, we measured the temperature dependence of the quantum oscillations amplitude for samples of different Tl concentrations, with the field oriented along or close to the [100] direction. For a sample of $x=0.3\%$, this was done by measuring magnetoresistance curves up to a DC field of 30 T. For samples with compositions $x=0.15\%$ and $x=0.8\%$ shifts in the resonance frequency of a proximity detector oscillator (PDO) circuit were recorded as a function of field \cite{PDOref}, from which changes in the skin depth with field, and therefore, resistivity, can be obtained \cite{PDOYBCO}. To accurately determine the cyclotron effective masses for such a low carrier-density material, we simultaneously fitted all magnetoresistance/PDO frequency shift curves to the Lifshitz-Kosevich (LK) formula (in SI units) \cite{Shoenberg}:

\begin{eqnarray}
\frac{\rho(H) -\rho_{0}}{\rho_0} & & =\sum_{i} C_i \left\{\exp\left(\frac{-\num{14.7}(m_i^{cyc}/m_e) \Theta_{D,i})}{H}\right)\right\} \nonumber \\
& & \times\left\{\frac{T/H}{\sinh\left(\num{14.7}(m_i^{cyc}/m_e)T/H\right)}\right\} \nonumber \\
& & \times\cos\left[2\pi\frac{f_i}{H}+\phi_i\right] \label{eq_LKmass}
\end{eqnarray}

where the sum is over each frequency observed in the data, and for which a separate cyclotron effective mass, $m_i^{cyc}/m_e$ can be obtained.

Figure \ref{fig_massTl} shows the temperature dependence of the oscillating component of magnetoresistance/PDO frequency shift for the Pb$_{1-x}$Tl$_x$Te samples studied, with field oriented along the [100] direction, providing direct access to the cyclotron effective mass along the [100] direction, $m_{[100]}^{cyc}$. Least-squares fits to Eq. \ref{eq_LKmass}, including the two strongest frequency components for each Tl doping, and for a field range of 7 T to 45 T (pulsed fields, at Los Alamos National Lab) for the $x=0.15\%$ sample, 7 T to 30 T (DC field) for the $x=0.3\%$ sample, and 17 T to 59 T (pulsed fields) for the $x=0.8\%$ sample, are shown in the right-column plots of this figure. The $m_{[100]}^{cyc}$ cyclotron mass obtained from these fits are shown in the plots, and summarized in Table \ref{table_Tlparam}. Figure \ref{fig_masspLutt} shows the obtained values of cyclotron effective masses along the [100] direction as a function of Fermi energy, for the Tl-doped samples studied, as well as for the Na-doped samples reported in ref. \citenum{GiraldoNaPRB}. An enhancement on the cyclotron effective mass for Fermi energies above the critical one is observed, which suggests an enhancement in electronic correlations in the L-pockets for compositions $x>x_c=0.3\%$. 
Since the mass renormalization appears to turn on for $x>x_c$, it is very likely connected to the presence of the resonant Tl impurity states at the Fermi energy. Whether this reflects an enhanced electron-phonon coupling associated with these states \cite{KotliarPRX2013}, the presence of a Suhl resonance from the putative charge Kondo effect \cite{DZero,Nagaoka,Suhl}, or some other interaction deriving from the impurity states, remains to be established.

Within the Kane model (known to be valid for PbTe in the absence of impurity states \cite{GiraldoNaPRB}) the DoS depends on the cyclotron effective mass like $g(E)\propto m_d^{3/2}$, where $m_d=[m^{cyc}_{\|}(m^{cyc}_{\bot})^2]^{1/3}$, and $m^{cyc}_{\|}$ and $m^{cyc}_{\bot}$ are the longitudinal and transverse cyclotron masses for the ellipsoidal L-pockets (for a perfect ellipsoid of anisotropy $K=14.3$, $m_d\approx 0.92*m_{[100]}^{cyc}$). The present case is however more consistent with a renormalisation of the effective mass due to interactions which implies a linear dependence of the DoS on effective mass. As these scenarios give an increase in the DoS of 1.7 and 1.4 respectively for a sample with $x=0.8\%$, they cannot explain the observed missing density of states, which is found to be 3.6 higher from the Sommerfeld coefficient than that derived from the L-pocket via the Kane model, as illustrated in figure 4(c) of the main text.

\end{document}